\documentclass[10pt,twocolumn,english]{revtex4-1}
\usepackage[T1]{fontenc}
\usepackage[latin9]{inputenc}
\usepackage{amsmath}
\usepackage{graphicx}
\usepackage{esint}

\makeatletter
\usepackage{babel}

\usepackage{babel}

\makeatother

\usepackage{babel}
\begin{document}

\title{Entanglement of  a two-atom system driven  by the quantum vacuum 
in arbitrary cavity size}

\author{G. Flores-Hidalgo }

\email{gfloreshidalgo@unifei.edu.br}

\affiliation{Instituto de F\'{\i}sica e Qu\'{\i}mica, Universidade Federal de Itajub\'a, 
37500-903, Itajub\'a, MG, Brazil.}

\author{M. Rojas}

\email{moises.leyva@dfi.ufla.br}

\author{Onofre Rojas}

\email{ors@dfi.ufla.br}

\affiliation{Departamento de F\'{\i}sica, Universidade Federal de Lavras, CP 3037,
37200-000, Lavras, MG, Brazil.}
\begin{abstract}
We study the dynamical entanglement of two identical atoms interacting with
a quantum  field. As a simplified model for this physical system  we consider two 
harmonic oscillators linearly coupled to a massless scalar field in the dressed 
coordinates and states approach and enclose the
whole system inside a spherical cavity of radius $R$.  
Through a quantity
called concurrence, the entanglement evolution for the two-atom system will be discussed,
for a range of initial states composed of a superposition of atomic
states. Our results reveals how the concurrence of the two atoms behaves
through the time evolution, for arbitrary cavity size and for arbitrary
coupling constant, weak, intermediate or strong. All our computations are exact and only the final
step is numerical.
These numerical solutions give us fascinating results for the concurrence, such as quasi-random
fluctuations, with a resemblance of periodicity. Another
interesting result we found is when the system is initially maximally
entangled (disentangled), after the time $t=2R$, the system becomes again strongly entangled
(disentangled)  particularly during the first oscillations,
later this phenomenon could be wrecked depending on the initial condition.
 We also show the
concurrence after a too long time elapsed with a good precision. 
\\
{\bf PACS}: 03.67.Mn,~03.67.Bg
\pacs{03.67.Mn,~03.67.Bg}

\end{abstract}
\maketitle

\section{Introduction}

Entanglement is one of the most exciting topics in quantum mechanics,
due it's unusual property of non-locality, which has been considering
as a key source in quantum information processing. During the past
decade, due the rapid development of the experimental process in quantum
control, there has been a rapidly growing interest in entanglement
generation. There are potential applications in nearly all quantum
communication and computation protocols \cite{nielsen}, such as quantum
teleportation \cite{Bennett,Bouwmeester}, quantum secure direct communication \cite{Marzolino},
quantum computation \cite{Bennett-1,Deng} among others.

One of the simplest schemes in which entanglement can be generated
is a system containing a couple of two-level atoms. In 90 decade,
two-atom entangled states have already been demonstrated experimentally
using ultra cold trap ions \cite{DeVoe,Turchette} and cavity quantum
electrodynamics (QED) schemes \cite{Hagley}. It has been showed that
entangled states can be generated in a two-atom system by a continuous
driving of the atoms with a coherent or chaotic thermal field \cite{Schneider,Kim},
or by a pulse excitation followed by a continuous radiative decay \cite{Plenio,Beige,Cabrillo}.
In order to confront the experimental results, several theoretical
studies have been considered for the preparation of a two-atom system
in an entangled state. In references \cite{Ficek,Tana} the entanglement
of two atoms by spontaneous emission for a model that includes dipole-dipole
interaction between the atoms have been discussed. The authors studied
both numerically and analytically the concurrence and negativity for
such a system. Also, in reference \cite{Tana-1} the authors discussed
numerical solutions for the concurrence in a system of two atoms driven
by a laser field.

In this work, we study the entanglement of a two-atom system driven
by the quantum vacuum field, the whole system enclosed inside a spherical cavity of radius
$R$. As simplified model of the atoms we consider each one as single
harmonic oscillator and the radiation field is taken as a massless
scalar field. In order to give physical sense to our model we will
work in the dressed coordinates and dressed states approach, that
were introduced some time ago in references \cite{adolfo1,adolfo2,gabriel},
in analogy with the renormalized fields in quantum field theory. Such
dressed coordinates and states have been defined in the physical context
of an atom, approximated by an harmonic oscillator, linearly coupled
to a scalar field and confined in a spherical
cavity of radius $R$. In terms of dressed coordinates, dressed states
have been defined as the physically measurable states. The dressed
states having the physical correct property of stability of the oscillator
(atom) ground state in the absence of field quanta (the quantum vacuum).
For a clear explanation see reference \cite{yony}. Also, the formalism
showed to have the technical advantage of allowing  exact computation
of the probabilities associated with the different oscillator (atom)
radiation processes \cite{casana}. For example, it has been obtained
easily the probability of the atom to decay spontaneously from the
first excited state to the ground state for arbitrary coupling constant,
weak or strong and for arbitrary cavity size. For weak coupling constant
in the limit $R\to\infty$, that corresponds to free space, the old know result: $e^{-\Gamma t}$,
has been obtained \cite{adolfo1}. Also, considering a cavity of sufficiently
small radius \cite{adolfo2}, the method accounted for the experimentally
observed inhibition of the spontaneous decaying processes of the
atom \cite{hulet,haroche2}. In references \cite{nonlinear,yony} the concept
of dressed coordinates and states have been extended to the case in
which nonlinear interactions between the oscillator and the field
modes are taken into account. Furthermore, in \cite{eletromag}
the oscillator electromagnetic field interaction has been considered
and in reference \cite{casana2} dressed coordinates and states have been
introduced in the path integral formalism. 

This work is organized as follow. In section II we describe the model
of the two atoms coupled to a massless scalar field,
find the eigenvalues and eigenfunctions of the system and define the
dressed coordinates and states. In section III we compute some probability
amplitudes, that will be useful for further study of the entanglement
of the two-atom system inside the spherical cavity. In section IV we discuss the
 entanglement  of the
two-atom system through  the quantitity called of  concurrence,  illustrating  the
behaviour of this quantity as a funtion of time, coupling constant and cavity radius.
Finally, in section V we give our conclusions.  We consider natural units $\hbar=c=1$.

\section{The Model}

Let us consider two identical atoms inside a cavity linearly coupled
to a massless scalar field. Roughly approximating the atoms by two
harmonic oscillators of frequency $\omega_{0}$, the system can be
described by the following Hamiltonian, 
\begin{alignat}{1}
H= & \tfrac{1}{2}(p_{A}^{2}+\omega_{0}^{2}q_{A}^{2})+\tfrac{1}{2}(p_{B}^{2}+
\omega_{0}^{2}q_{B}^{2})+\tfrac{1}{2}\sum_{k=1}^{N}(p_{k}^{2}+\omega_{k}^{2}q_{k}^{2})+\nonumber \\
 & +\tfrac{1}{\sqrt{2}}\sum_{k=1}^{N}c_{k}q_{k}(q_{A}+q_{B})+
\sum_{k=1}^{N}\frac{c_{k}^{2}}{4\omega_{k}}(q_{A}+q_{B})^{2},\label{hamiltonian}
\end{alignat}
where the coordinates and momenta $q_{A},p_{A}$ and $q_{B},p_{B}$
correspond to the atoms $A$ and $B$ respectively, $\omega_{k}$ is the
frequency of the field modes and $c_{k}$ the coupling constant of
the atoms with the scalar field . The coordinates and momenta $q_{k}$
and $p_{k}$ are related to the field modes of frequencies $\omega_{k}$.
The last term in Eq. (\ref{hamiltonian}) is introduced to guarantee the
positiveness of the Hamiltonian and can be understood as a renormalization
frequency for the particles oscillators \cite{weiss}. At the end of
the calculation, we will take $N\to\infty$. Also $\omega_{k}=\pi k/R$,
$k=\{0,1,2,...\}$, where $R$, is
the radius in which the whole system is confined. Surely, we can recover
the free  space case taking the limit $R\to\infty$.

The model Hamiltonian given by Eq. (\ref{hamiltonian}) has
been introduced in Ref. \cite{hu} and used to study the entanglement
between two single harmonics oscillators in Ref. \cite{paz}. Here
we re-derive above Hamiltonian by considering two identical harmonic
oscillator located at positions ${\bf r}_{A}$ and ${\bf r}_{B}$,
interacting with a massless scalar field, the whole system inside
a spherical cavity or radius $R$. The Lagrangian of the system is
\begin{alignat}{1}
L= & \frac{1}{2}(\dot{q}_{A}^{2}-\omega_{0}^{2}q_{A}^{2})+\frac{1}{2}(\dot{q}_{B}^{2}-\omega_{0}^{2}q_{B}^{2})+\nonumber \\
 & +\frac{1}{2}\int d^{3}{\bf r}\left(\left(\tfrac{\partial\phi}{\partial t}\right)^{2}-({\bf \nabla}\phi)^{2}\right)+\nonumber \\
 & -\sqrt{g}\negmedspace\int\negmedspace d^{3}{\bf r}\phi({\bf r},t)\Big(\!\delta({\bf r}-{\bf r}_{A})q_{A}\!+\!\delta({\bf r}-{\bf r}_{B})q_{B}\!\Big),\label{lagrangean}
\end{alignat}
where the coupling constant is written in such a way that $g$ has
dimension of frequency. Note that the two harmonic oscillators oscillate
in another abstract space than spatial $\Re^{3}$. The coordinates
$q_{A}$ and $q_{B}$ must be viewed as coordinates representing all
the relevant internal degrees of freedom of the subsystems $A$ and
$B$. We are also considering unit mass for the particle oscillators.
For other masses the Lagrangian can be transformed into Eq. (\ref{lagrangean}),
through a scale coordinate transformation. Now, suppose that the particle
oscillators (atoms) are located nearly at the center of the cavity,
this means that ${\bf r}_{A}\approx{\bf r}_{B}\approx0$. Therefore,
the Lagrangian becomes 
\begin{eqnarray}
L & = & \frac{1}{2}(\dot{q}_{A}^{2}-\omega_{0}^{2}q_{A}^{2})+\frac{1}{2}(\dot{q}_{B}^{2}-\omega_{0}^{2}q_{B}^{2})+\nonumber \\
 &  & +\frac{1}{2}\int d^{3}{\bf r}\left(\left(\tfrac{\partial\phi}{\partial t}\right)^{2}-({\bf \nabla}\phi)^{2}\right)+\nonumber \\
 &  & -\sqrt{g}\:\phi({\bf 0},t)\left(q_{A}+q_{B}\right).\label{lagrangean1}
\end{eqnarray}
Expanding the field $\phi({\bf r},t)$  as 
\begin{equation}
\phi({\bf r},t)=\sum_{k}q_{k}(t)u_{k}({\bf r})\;,\label{expansion}
\end{equation}
where $\{u_{k}({\bf r})\}$ is a complete orthonormal set of solutions
of the equation, 
\begin{equation}
-\nabla^{2}u_{k}({\bf r})=\omega_{k}^{2}u_{k}({\bf r}),~~~u_{k}(R)=0\;,\label{operator}
\end{equation}
where we are imposing Dirichlet boundary conditions for the field,
$\phi(R,t)=0$. Replacing (\ref{expansion}) in (\ref{lagrangean1}),
using Eq. (\ref{operator}) and orthonormality property, we get 
\begin{eqnarray}
L & = & \frac{1}{2}(\dot{q}_{A}^{2}-\omega_{0}^{2}q_{A}^{2})+\frac{1}{2}(\dot{q}_{B}^{2}-\omega_{0}^{2}q_{B}^{2})+\frac{1}{2}\sum_{k}\left(\dot{q}_{k}^{2}-\omega_{k}^{2}q_{k}^{2}\right)\nonumber \\
 &  & -\sqrt{g}\sum_{k}u_{k}(0)q_{k}\left(q_{A}+q_{B}\right)\;.\label{lagrangean2}
\end{eqnarray}
From above equation we can note that the field modes interacting with
the particle oscillators are such that $u_{k}(0)\neq0$. Solving Eq. (\ref{operator}), it can 
be showed that such solutions are the spherically
symmetric ones, \textit{i.e}, 
\begin{equation}
u_{k}({\bf r})=\frac{\sin(\omega_{k}r)}{r\sqrt{2\pi R}},\label{eigenfunctions}
\end{equation}
where 
\begin{equation}
\omega_{k}=\frac{\pi k}{R},~k=1,2,3,...,N\to\infty.
\end{equation}
Using Eq. (\ref{eigenfunctions}), we get 
\begin{equation}
u_{k}(0)=\frac{\omega_{k}}{\sqrt{2\pi R}}\;.\label{zerovalued}
\end{equation}
Substituting Eq. (\ref{zerovalued}) in Eq. (\ref{lagrangean2}),
we obtain 
\begin{eqnarray}
L & = & \frac{1}{2}(\dot{q}_{A}^{2}-\omega_{0}^{2}q_{A}^{2})+\frac{1}{2}(\dot{q}_{B}^{2}-\omega_{0}^{2}q_{B}^{2})+\frac{1}{2}\sum_{k}\left(\dot{q}_{k}^{2}-\omega_{k}^{2}q_{k}^{2}\right)\nonumber \\
 &  & -\sqrt{g}\sum_{k}u_{k}(0)q_{k}\left(q_{A}+q_{B}\right).
\end{eqnarray}
From which, it is not difficult to obtain the Hamiltonian (\ref{hamiltonian}),
with $c_{k}$ given by 
\begin{equation}
c_{k}=\frac{\omega_{k}}{\pi\sqrt{2}}\sqrt{g\Delta\omega_{k}},~~~\Delta\omega_{k}=\frac{\pi }{R}.
\end{equation}

In order to diagonalize the Hamiltonian (\ref{hamiltonian}), we 
introduce the relative and the center of mass  coordinates
$q_{-}$ and $q_{0}$, respectively and given by 
\begin{equation}
q_{-}=\tfrac{1}{\sqrt{2}}(q_{A}-q_{B}),~~~q_{0}=\tfrac{1}{\sqrt{2}}(q_{A}+q_{B}).\label{center}
\end{equation}
Substituting the above relations in Eq. (\ref{hamiltonian}), we have
\begin{alignat}{1}
H= & \tfrac{1}{2}(p_{-}^{2}+\omega_{0}^{2}q_{-}^{2})+\tfrac{1}{2}(p_{0}^{2}+\omega_{0}^{2}q_{0}^{2})+\tfrac{1}{2}\sum_{k=1}^{N}(p_{k}^{2}+\omega_{k}^{2}q_{k}^{2})+\nonumber \\
 & +\sum_{k=1}^{N}c_{k}q_{k}q_{0}+\sum_{k=1}^{N}\frac{c_{k}^{2}}{2\omega_{k}}q_{0}^{2}.\label{ham1}
\end{alignat}
Note that in the Hamiltonian \eqref{ham1}, the relative coordinate
$q_{-}$ is decoupled, thus, one can diagonalize the Hamiltonian ignoring
the first term. For this purpose, we introduce, the collective coordinates
and momenta, $Q_r$ and $P_r$, given by 
\begin{equation}
q_{\mu}=\sum_{r}t_{\mu}^{r}Q_{r},~~~p_{\mu}=\sum_{r}t_{\mu}^{r}P_{r},\label{diag1}
\end{equation}
where $\mu=0,k$, and $t_{\mu}^{r}$ is given by

\begin{equation}
t_{k}^{r}=\tfrac{c_{k}}{\omega_{k}^{2}-\Omega_{r}^{2}}t_{0}^{r},~~t_{0}^{r}=\left(1+\sum_{k=1}^{N}\tfrac{c_{k}^{2}}{(\omega_{k}^{2}-\Omega_{r}^{2})^{2}}\right)^{-\frac{1}{2}},\label{matrix}
\end{equation}
It is worth to mention that the matrix $\{t_{\mu}^{r}\}$ is orthogonal
and satisfy the following relations 
\begin{equation}
\sum_{\mu}t_{\mu}^{r}t_{\mu}^{s}=\delta_{rs},~\text{and}~~\sum_{r}t_{\mu}^{r}t_{\nu}^{r}=\delta_{\mu\nu}.\label{ortog}
\end{equation}
The Hamiltonian (\ref{ham1}) can be rewritten in collective coordinates,
which simply reduce to 
\begin{equation}
H=\tfrac{1}{2}(p_{-}^{2}+\omega_{0}^{2}q_{-}^{2})+\tfrac{1}{2}\sum_{r}(P_{r}^{2}+\Omega_{r}^{2}Q_{r}^{2}),\label{diag2}
\end{equation}
where the normal frequencies $\Omega_{r}$ are the solutions of the
equation 
\begin{equation}
\omega_{0}^{2}-\Omega_{r}^{2}=\sum_{k=1}^{N}\frac{c_{k}^{2}\Omega_{r}^{2}}{\omega_{k}^{2}(\omega_{k}^{2}-\Omega_{r}^{2})}.\label{normalf}
\end{equation}
Now, we are ready to write the eigenfunctions and energy eigenvalues
of the system. The eigenfunctions are given by 
\begin{equation}
\phi_{n_{-},n_{0},n_{1},..}(q_{-},Q)=\phi_{n_{-}}(q_{-})\prod_{r=0}\phi_{n_{r}}(Q_{r}),\label{eigen}
\end{equation}
where $\phi_{n_{-}}(q_{-})$, $\phi_{n_{r}}(Q_{r})$ are  one dimensional
harmonic oscillator eigenfunctions of frequencies $\omega_{0}$
and $\Omega_{r}$ respectively. Whereas the corresponding energy
eigenvalues are, 
\begin{equation}
E_{n_{-},n_{0},n_{1},...}=\left(n_{-}+\tfrac{1}{2}\right)\omega_{0}+\sum_{r=0}\left(n_{r}+\tfrac{1}{2}\right)\Omega_{r},\label{ener}
\end{equation}
where $n_{-},n_{r}=\{0,1,2,...\}$. 

\subsection{Dressed coordinates and states}
In what follows, we
introduce the dressed coordinates $q_{A}'$, $q_{B}'$ and $q_{k}'$,
for the atoms $A$, $B$ and field modes respectively. In terms of
this coordinates, we define the dressed states as follows
\begin{equation}
\psi_{n_{A},n_{B},n_{k}}(q')=\psi_{n_{A}}(q_{A}')\psi_{n_{B}}(q_{B}')\prod_{k=1}\psi_{n_{k}}(q'_{k})\,,\label{dressed}
\end{equation}
where $\psi_{n_{A}}(q_{A}')$, $\psi_{n_{B}}(q_{B}')$ and $\psi_{n_{k}}(q'_{k})$
are one dimensional harmonic oscillators eigenfunctions, with
frequencies $\omega_{A}=\omega_{0}$, $\omega_{B}=\omega_{0}$ and
$\omega_{k}$, respectively. We define such dressed states as the
physically realizable states. The dressed state as given by Eq.
(\ref{dressed}) represent the state in which the atom $A$ is in the energy level $n_A$, the atom $B$ is in the
energy level $n_B$ and there are $n_k$ field quanta of frequencies $\omega_k$. 
In general, the dressed states defined above are unstable states, because
interaction, such states  could decay into other measurable
states. For example, the state $\psi_{1,0,0,..,0}(q')$ in which the atom $A$ is in its first excited
level could decay to its ground state by emission of field quanta or due to the absortion of
energy by the atom $B$. In this case, the final possible states are the dressed ones
$\psi_{0,0,..0,1_k,0,..0}(q')$ or $\psi_{0,1,0,...,0}(q')$. 

On the other hand,  the dresed ground state $\psi_{0,0,0...}(q')$ that describes the
two atoms in the ground state and no field quanta,  must be stable, according experimental
facts. The last one allow us to give an analytical expression for
the dressed coordinates $q_{A}'$, $q_{B}'$ and $q_{k}'$ in terms
of collective coordinates or in terms of the bare ones. 
The stability
requires that the dressed ground state $ \psi_{0,0,...,0}(q')$
 must be one of the energy eigenfunctions
of the Hamiltonian (\ref{diag2}), and it is natural to associate
it with the ground state of the system, $\phi_{0,0,...,0}(q_{-},Q)$. Thus, we can obtain the dressed coordinates
from
\begin{equation}
\psi_{0,0,...,0}(q')\propto\phi_{0,0,...,0}(q_{-},Q).
\label{defdre}
\end{equation}
The ground state of the system and the dressed ground state are given respectively by
\begin{equation}
\phi_{0,0,...,0}(q_{-},Q)\propto{\rm e}^{-\tfrac{1}{2}\left(\omega_{0}q_{-}^{2}+\underset{r}{\sum}\Omega_{r}Q_{r}^{2}\right)}.\label{groundc}
\end{equation}
and
\begin{equation}
\psi_{0,0,...,0}(q')\propto{\rm e}^{-\frac{1}{2}\left(\omega_{0}q'_{A}\!\!\!~^{2}+\omega_{0}q'_{B}\!\!\!~^{2}+\underset{k=1}{\sum}\omega_{k}q'_{k}\!\!\!~^{2}\right)}.\label{dressc}
\end{equation}
In order to find the relation between dressed and collective coordinates,
 let us define the dressed relative coordinate $q'_{-}$ and dressed center of mass
coordinate $q_{0}'$, in the same way as for the bare ones
(\ref{center}), we have 
\begin{equation}
q'_{-}=\tfrac{1}{\sqrt{2}}(q'_{A}-q'_{B}),~~~q'_{0}=\tfrac{1}{\sqrt{2}}(q'_{A}+q'_{B}).\label{drescen}
\end{equation}
Substituting in terms of relative and center of mass dressed coordinates,
the dressed ground state (\ref{dressc}), becomes, 
\begin{equation}
\psi_{0,0,...,0}(q')\propto{\rm e}^{-\frac{1}{2}\left(\omega_{0}q'_{-}\!\!\!~^{2}+\omega_{0}q'_{0}\!\!\!~^{2}+\underset{k=1}{\sum}\omega_{k}q'_{k}\!\!\!~^{2}\right)}.\label{dresscb}
\end{equation}
Using Eqs. (\ref{groundc}) and (\ref{dresscb}) in (\ref{defdre}) we have, 
\begin{equation}
q'_{-}=q_{-},~~~q'_{\mu}=\sum_{r=0}\sqrt{\tfrac{\Omega_{r}}{\omega_{\mu}}}\:t_{\mu}^{r}Q_{r},\quad\mu=0,k,\label{dreco}
\end{equation}
where we have considered the property given by Eq. (\ref{ortog}).
Finally, using Eq. (\ref{drescen}) in (\ref{dreco}), we are able
to write the dressed coordinates for the two atoms and field
modes, 
\begin{eqnarray}
q'_{A} & = & \tfrac{1}{\sqrt{2}}\left(\sum_{r=0}\sqrt{\tfrac{\Omega_{r}}{\omega_{0}}}\:t_{0}^{r}Q_{r}+q_{-}\right),\label{atomA}\\
q'_{B} & = & \tfrac{1}{\sqrt{2}}\left(\sum_{r=0}\sqrt{\tfrac{\Omega_{r}}{\omega_{0}}}\:t_{0}^{r}Q_{r}-q_{-}\right),\label{atomB}\\
q'_{k} & = & \sum_{r=0}\sqrt{\tfrac{\Omega_{r}}{\omega_{k}}}t_k^r\:Q_{r}.\label{field}
\end{eqnarray}
As will be shown in the following section, the above relationships
will be of fundamental value to perform quantum mechanical calculations.

\section{Probability amplitudes}

In this section, we obtain the different probability amplitudes between
the dressed states introduced in the last section. These quantities
will give us the probability of transitions between the different
atom-atom-field states. Suppose that at time $t=0$, the state
of the system is given by $\psi_{n_{A},n_{B},n_{1},...}(q')$, where
 the atom $A$ is in the energy level $n_{A}$, the atom $B$ is
in the energy level $n_{B}$ and there are $n_{k}$ field quanta of
frequencies $\omega_{k}$, $k=\{1,2,...\}$. We want to obtain the
probability amplitude to find the system at time $t$ in the state
$\psi_{m_{A},m_{B},m_{1},...}(q')$ , \textit{i.e}, the state in which
the atom $A$ is in the energy level $m_{A}$, the atom $B$ is in
the energy level $m_{B}$, and there are $m_{k}$ field quanta of
frequencies $\omega_{k}$. Denoting such probability amplitude by
${\cal A}_{n_{A},n_{B},n_{1},...}^{m_{A},m_{B},m_{1},...}(t)$, we
have, 
\begin{equation}
{\cal A}_{n_{A},n_{B},n_{1},...}^{m_{A}\!,\!m_{B}\!,\!m_{1}\!,...}(t)\!=
\!{}_{d}\langle m_{A}\!,\!m_{B}\!,\!m_{1}\!,...|\mathrm{e}^{-i\hat{H}t}|n_{A}\!,\!n_{B}\!,\!n_{1}\!,...\rangle_{d},
\end{equation}
where the dressed kets $|n_{A},n_{B},n_{1},...\rangle_{d}$ are such
that, $\psi_{n_{A},n_{B},n_{1},...}(q')=\langle q'_{A},q'_{B},q'_{1},...|n_{A},n_{B},n_{1},...\rangle_{d}$.

For further use, we will focus only on the probability amplitudes
between low-level energy atomic states. Let us consider, for example,
that at time $t=0$ the atom $A$ is in the first excited level, atom
$B$ is on the ground state and there are no field quanta. In this
case the state of the system is given by

\begin{equation}
\psi_{1,0,0,...,0}(q')=\psi_{0,0,0,..}(q')H_{1}\left(\sqrt{\omega_{0}}q_{A}'\right)\,.\label{atomfirst}
\end{equation}
Rewriting the above eigenfunction, in terms of collective coordinates,
using $\psi_{0,0,...,0}(q')=\phi_{0,0,...,0}(q_{-},Q)$, Eqs. (\ref{atomA})
and (\ref{eigen}), we have 
\begin{equation}
\psi_{1,0,0,..,0}(q')=
\frac{1}{\sqrt{2}}\Big(\phi_{1,0,0,...,0}+\underset{r=0}{\sum}t_{0}^{r}\phi_{0,..,0,1_{r},0,..,0}\Big)\,.\label{expand}
\end{equation}
At time $t$,  this state will evolve to, 
\begin{alignat}{1}
\psi_{1,0,0,..,0}(q',t)= & \frac{\mathrm{e}^{-iE_{0}t}}{\sqrt{2}}\Big(\mathrm{e}^{-i\omega_{0}t}\phi_{1,0,0,..,0}+
\nonumber \\
 & +\sum_{r=0}t_{0}^{r}\mathrm{e}^{-i\Omega_{r}t}\phi_{0,..,0,1_{r},0,..,0}\Big),\label{evolved}
\end{alignat}
with $E_{0}=(\omega_{0}+\sum_{r}\Omega_{r})/2$, the ground
state energy of the total system.

Now, we can easily calculate the different probability amplitudes
associated with the different possible transitions. We can perform
the probability amplitudes with these different possibilities as fallows.
The probability amplitude to find the state $\psi_{1,0,0,...,0}$, to
remain in the initial state, is obtained by 
\begin{eqnarray}
{\cal A}_{1,0,0,...}^{1,0,0,..}(t) & = & \int dQdq_{-}\psi_{1,0,0,...,0}^{\ast}(q')\psi_{1,0,0...,0}(q',t)\nonumber \\
 & = & \frac{\mathrm{e}^{-iE_{0}t}}{2}\left(\mathrm{e}^{-i\omega_{0}t}+f_{00}(t)\right)\label{ampli1}
\end{eqnarray}
where, 
\begin{equation}
f_{00}(t)=\sum_{r=0}(t_{0}^{r})^{2}\mathrm{e}^{-i\Omega_{r}t}\,.\label{foo}
\end{equation}
In order to compute the probability amplitude that the atom $A$ decays
into its ground state and the atom $B$ jump from its ground state
to its first excited level, we write such final possible state, $\psi_{0,1,0,...,0}(q')$,
in terms of collective coordinates, 
\begin{equation}
\psi_{0,1,0,..,0}(q')=\frac{1}{\sqrt{2}}\Big(\underset{r=0}{\sum}t_{0}^{r}\phi_{0,..,0,1_{r},0,..,0}-\phi_{1,0,0,..,0}\Big),
\end{equation}
and taking the scalar product with (\ref{evolved}) we have, 
\begin{equation}
{\cal A}_{1,0,0,...}^{0,1,0,...}(t)=\frac{\mathrm{e}^{-iE_{0}t}}{2}\left(f_{00}(t)-\mathrm{e}^{-i\omega_{0}t}\right).\label{ampli2}
\end{equation}
In a similar way, considering $\psi_{0,...,0,1_{k},0,...,0}(q')$, the
state in which the two atoms are in the ground state and there is a
field quanta of frequency $\omega_{k}$, could be expressed it in
terms of collective coordinates as follow, 
\begin{equation}
\psi_{0,..,0,1_{k},0,...,0}(q')=\sum_{r=0}t_{k}^{r}\phi_{0,...,0,1_{r},0...}\label{1field}
\end{equation}
and taking the scalar product with (\ref{evolved}) we find the probability
amplitude for atom $A$ to decay from its first excited level by
emission of a field quanta of frequency $\omega_{k}$, 
\begin{equation}
{\cal A}_{1,0,0,...}^{0,0,..,0,1_{k},0,..}(t)=\frac{\mathrm{e}^{-iE_{0}t}}{\sqrt{2}}f_{0k}(t)\,,\label{ampli3}
\end{equation}
where 
\begin{equation}
f_{0k}(t)=\sum_{r=0}t_{0}^{r}t_{k}^{r}\mathrm{e}^{-i\Omega_{r}t}\,.\label{fok}
\end{equation}
The states $\psi_{1,0,0,...,0}(q')$, $\psi_{0,1,0,...,0}(q')$ and $\psi_{0,...,0,1_{k},0,...,0}(q')$
are the only possible ones in which the initial state $\psi_{1,0,0,...,0}(q')$
can be found at time $t$. To show this one, we take the probabilities
$|{\cal A}_{1,0,0,...}^{1,0,0,...}(t)|^{2}$, $|{\cal A}_{1,0,0,...}^{0,1,0,...}(t)|^{2}$
and $|{\cal A}_{1,0,0,...}^{0,..,0,1_{k},0,..}(t)|^{2}$, using the
eqs. (\ref{ampli1}), (\ref{ampli2}), (\ref{ampli3}) and (\ref{ortog}),
we find the following relation 
\begin{equation}
|{\cal A}_{1,0,0,...}^{1,0,0,...}(t)|^{2}+|{\cal A}_{1,0,0,...}^{0,1,0,...}(t)|^{2}+\sum_{k=1}|{\cal A}_{1,0,0,...}^{0,..,0,1_{k},0,..}(t)|^{2}=1,\label{sum1}
\end{equation}
that shows our assertion. Note that in (\ref{sum1}), the summation
was performed over all possible frequencies since the emitted field
quanta could be of any arbitrary frequency.

Now, considering as the initial state $\psi_{0,1,0,,...,0}(q')$, the
state in which the atom $B$ is in its first excited level, while
the atom $A$ is in its ground state, and there are no field quanta,
we can find, in similar way as above, all the probability amplitudes
related to such initial state. Because atoms $A$ and $B$ are identical,
the state of the system can be found at time $t$ in one of the states,
$\psi_{1,0,0,...,0}(q')$, $\psi_{0,1,0,...,0}(q')$ or $\psi_{0,..,0,1_{k},0,...,0}(q')$.
We find respectively for each one of these possible states, 
\begin{eqnarray}
{\cal A}_{0,1,0,...}^{1,0,0,...}(t) & = & {\cal A}_{1,0,0,...}^{0,1,0...}(t)\,,\nonumber \\
{\cal A}_{0,1,0,...}^{0,1,0,...}(t) & = & {\cal A}_{1,0,0,...}^{1,0,0...}(t)\,,\nonumber \\
{\cal A}_{0,1,0,...}^{0,...,0,1_{k},0...}(t) & = & {\cal A}_{1,0,0,...}^{0,...,0,1_{k},0...}(t)\,.\label{ampli456}
\end{eqnarray}

In order to shorten the notation for the dressed states, $\psi_{1,0,0,...,0}(q')$,
$\psi_{0,1,0,...,0}(q')$ and $\psi_{0,0,..,0,1_{k},0,...,0}(q')$, we
can write the dressed states using Dirac notation, respectively by
$|1,0,0,...\rangle_{d}$, $|0,1,0,...\rangle_{d}$ and $|0,0,...,0,1_{k},0,...\rangle_{d}$.
Now, using the results obtained above, we can write the dressed states
$|1,0,0,...\rangle_{d}$ and $|0,1,0,...\rangle_{d}$, at time $t$,
respectively as 
\begin{alignat}{1}
|1,0,0,...;t\rangle_{d}= & \sum_{k=1}{\cal A}_{1,0,0,...}^{0,...,0,1_{k},0...}(t)|0,0,...,0,1_{k},0,..\rangle_d+\nonumber \\
 & +{\cal A}_{1,0,0,...}^{1,0,0,...}(t)|1,0,0,...\rangle_{d}\nonumber \\
 & +{\cal A}_{1,0,0,...}^{0,1,0...}(t)|0,1,0,...\rangle_{d},\label{atomat}
\end{alignat}
and 
\begin{alignat}{1}
|0,1,0,...;t\rangle_{d}= & \sum_{k=1}{\cal A}_{0,1,0,...}^{0,...,0,1_{k},0...}(t)|0,0,...,0,1_{k},0,..\rangle_d\nonumber \\
 & +{\cal A}_{0,1,0,...}^{0,1,0,...}(t)|0,1,0,...\rangle_{d}\nonumber \\
 & +{\cal A}_{0,1,0,...}^{1,0,0...}(t)|1,0,0,...\rangle_{d}.\label{atombt}
\end{alignat}
In next section, when we consider the dynamics of the entanglement of specific
 atomic states, we will use the above time dependent dressed states.

From expressions given by Eqs. (\ref{ampli1}), (\ref{ampli2}) and
(\ref{ampli3}), we note that to obtain the time evolution of vector
states, as given by Eqs. (\ref{atomat}) and (\ref{atombt}), we have
to compute the coefficients $f_{\mu\nu}$, Eqs. (\ref{foo}) and (\ref{fok}).
It is a formidable task since before we have to compute the coefficients
$t_{\mu}^{r}$ and collective frequencies $\Omega_{r}$, given by
Eqs. (\ref{matrix}) and (\ref{normalf}). For limiting
cases of large (infinite) cavity and very small cavity one can obtain
analytical results for $f_{00}$ and $f_{0k}$. Nevertheless, for intermediate  cavity size, finding
$f_{00}$ and $f_{0k}$, becomes a cumbersome task. Then, for intermediate
cavity size we have resort to numerical computations. For this purpose
Eq. (\ref{normalf}) is expressed as \cite{adolfo1}, 
\begin{equation}
\cot(R\Omega_{r})=\frac{\Omega_{r}}{\pi g}+\frac{1}{R\Omega_{r}}\left(1-\frac{R\omega_{0}^{2}}{\pi g}\right).\label{eq:Omega}
\end{equation}
This transcendental equation can be solved numerically for the collective
frequencies $\Omega_{r}$ and using the numerical solutions we can
find $(t_{0}^{r})^{2}$, the term of Eq. \eqref{foo}, which can be
recast in the form, 
\begin{equation}
(t_{0}^{r})^{2}=\frac{\eta^{2}\Omega_{r}^{2}}{\left(\Omega_{r}^{2}-\omega_{0}\right)^{2}+\frac{\eta^{2}}{2}\left(3\Omega_{r}^{2}-\omega_{0}^{2}\right)+\pi^{2}g^{2}\Omega_{r}^{2}},
\end{equation}
where $\eta=\sqrt{2g\Delta \omega_k}$.
Finally, the infinite sum that appears in (\ref{foo}) or (\ref{fok})
can be done numerically, by noting that for large $\Omega_{r}$, the
coefficient $(t_{0}^{r})^{2}$ approaches zero as $1/(\Omega_{r})^{2}$.

\section{Entanglement of two-atom system  driven by quantum field}

Once we know the time evolution of vector states , we can construct
the time dependent reduced density operator for the two-atoms system
considered in last section and this will be useful to study
the dynamics of the entanglement   through the concurrence quantity.
Assuming the state of the system, at time $t=0$, is given by an arbitrary
linear combination of the states $|1,0,0,...\rangle_{d}$ and $|0,1,0,...\rangle_{d}$,
\begin{equation}
|\psi\rangle_{d}=\sqrt{\xi}\:|1,0,0,...\rangle_{d}+\sqrt{1-\xi}\:\mathrm{e}^{i\phi}|0,1,0,...\rangle_{d},\label{super}
\end{equation}
where in the first state the atom $A$ is in the first excited level
and in the second state, the atom $B$ is in the first excited level.
Whereas $\xi$ and $\phi$ are constant parameters that fix the initial
conditions. Using Eqs. (\ref{atomat}), (\ref{atombt}), and (\ref{ampli456})
we have the time dependent state $|\psi,t\rangle_{d}$, given by 
\begin{alignat}{1}
|\psi,t\rangle_{d}= & A(t)|1,0,0,...\rangle_{d}+B(t)|0,1,0,...\rangle_{d}+\nonumber \\
 & +\sum_{k=1}C_{k}(t)|0,0,...,0,1_{k},0,...\rangle_{d},\label{supert}
\end{alignat}
where 
\begin{eqnarray}
A(t) & = & \sqrt{\xi}{\cal A}_{1,0,0,...}^{1,0,0,...}(t)+\sqrt{1-\xi}\:\mathrm{e}^{i\phi}{\cal A}_{1,0,0,...}^{0,1,0,...}(t),\label{at}\\
B(t) & = & \sqrt{\xi}{\cal A}_{1,0,0,...}^{0,1,0,...}(t)+\sqrt{1-\xi}\:\mathrm{e}^{i\phi}{\cal A}_{1,0,0,...}^{1,0,0,...}(t),\label{bt}\\
C_{k}(t) & = & \left(\sqrt{\xi}+\sqrt{1-\xi}\mathrm{e}^{i\phi}\right){\cal A}_{1,0,0,...}^{0,0,...,0,1_{k},0...}.\label{ct}
\end{eqnarray}

The density operator for the state (\ref{supert}) is, 
\begin{equation}
\hat{\rho}(t)=|\psi,t\rangle_{d}~\!_{d}\langle\psi,t|\,.\label{mdensi}
\end{equation}
Tracing out over all field quanta degrees of freedom in (\ref{mdensi})
we have the reduced density matrix for the two-atom system, 
\begin{equation}
\hat{\rho}_{r}(t)=\sum_{n_{1},n_{2},...=0}~_{d}\langle n_{1},n_{2},...|\hat{\rho}(t)|n_{1},n_{2},...\rangle_{d}\,,\label{rdensi}
\end{equation}
where $|n_{1},n_{2},...\rangle_{d}$ are states restricted to the
sub-space of the degrees of freedom of the quantum field. Using (\ref{supert}),
the reduced density operator (\ref{rdensi}) becomes, 
\begin{alignat}{1}
\hat{\rho}_{r}(t)= & E(t)|0,0\rangle_{d}~\!_{d}\langle0,0|+\nonumber \\
 & +A(t)B^{\ast}(t)|1,0\rangle_{d}{}_{d}\langle0,1|+A^{\ast}(t)B(t)|0,1\rangle_{d}{}_{d}\langle1,0|+\nonumber \\
 & +|A(t)|^{2}|1,0\rangle_{d}{}_{d}\langle1,0|+|B(t)|^{2}|0,1\rangle_{d}{}_{d}\langle0,1|\label{rdensi1}
\end{alignat}
with $E(t)$ given by 
\begin{eqnarray}
E(t) & = & \sum_{k=1}|C_{k}(t)|^{2}\nonumber \\
 & = & 1-|A(t)|^{2}-|B(t)|^{2}\,.\label{constant}
\end{eqnarray}
where in passing to the second relation, we have used Eqs. (\ref{at})-(\ref{ct})
and (\ref{sum1}). In the basis $\{|0,0\rangle_{d},|1,0\rangle_{d},|0,1\rangle_{d},|1,1\rangle_{d}\}$,
the reduced density matrix is 
\begin{equation}
\rho_{r}(t)=\left(\begin{array}{cccc}
E(t) & 0 & 0 & 0\\
0 & |A(t)|^{2} & A(t)B^{\ast}(t) & 0\\
0 & A^{\ast}(t)B(t) & |B(t)|^{2} & 0\\
0 & 0 & 0 & 0
\end{array}\right).\label{matrixel}
\end{equation}

To measure the entanglement created by the two-atom system, we use the
concurrence as a measure of entanglement, which was introduced by
Wootters \cite{wootters}.
The concurrence can be obtained from the reduced density matrix $\rho_{r}$.
Considering the spin-flipped state, $\mathbf{R}(t)=
\rho_{r}(t)\sigma_{2}\otimes\sigma_{2}\rho_{r}(t)^{\ast}\sigma_{2}\otimes\sigma_{2}$,
and using (\ref{matrixel}), we have 
\begin{equation}
\mathbf{R}(t)=2\negthickspace\left(\negmedspace\begin{array}{cccc}
0 & 0 & 0 & 0\\
0 & |A(t)|^{2}|B(t)|^{2} & |A(t)|^{2}A(t)B^{\ast}(t) & 0\\
0 & |B(t)|^{2}A^{\ast}(t)B(t) & |A(t)|^{2}|B(t)|^{2} & 0\\
0 & 0 & 0 & 0
\end{array}\negmedspace\right).\label{concu}
\end{equation}
Therefore, the concurrence is defined as $\mathcal{C}(\rho_{r})=
{\rm max}\{0,\lambda_{1}-\lambda_{2}-\lambda_{3}-\lambda_{4}\}$,
where the $\lambda_{i}$'s are the square root of the eigenvalues
of matrix (\ref{concu}). After some algebraic manipulation, we obtain
\begin{equation}
\mathcal{C}(R,g,t,\xi,\phi)=2|A(t)B(t)|\,.\label{concu1}
\end{equation}

The concurrence, as given above, depends through $A(t)$ and $B(t)$
of frequency $\omega_{0}$, cavity radius $R$, coupling constant
with the scalar field $g$, initial conditions $\xi$ and $\phi$,
and certainly depends of time $t$. Some particular results can be
obtained, such as taking $\xi=1/2$ and $\phi=0$ in Eq. (\ref{super}). In this
case the initial state 
\begin{equation}
|\psi\rangle_d=\frac{1}{\sqrt{2}}\Big( |1,0,0,...\rangle_d+|0,1,0,...\rangle_d\Big)
\label{symmetric}
\end{equation}
is a symmetric maximally entangled state. Using Eqs. (\ref{ampli1}) and (\ref{ampli2}) in (\ref{at})-(\ref{bt}),
in this case the concurrence (\ref{concu1}) reduce to the following expression
\begin{equation}
\mathcal{C}(t,\xi=1/2,\phi=0)=|f_{00}(t)|^{2}\,,\label{dadoc1}
\end{equation}
which coincide with the probability
of a dressed atom to remain in its first excited level at time $t$
\cite{adolfo1}.

Another limiting case can be obtained when taking as initial state a maximally antisymmetric state,
corresponding to $\xi=1/2$ and $\phi=\pi$
in Eq. (\ref{super}),  the concurrence simply becomes 
\begin{equation}
\mathcal{C}(t,\xi=1/2,\phi=\pi)=1\,,\label{dadoc2}
\end{equation}
the two-atoms system is maximally entangled through the time, independent
of any cavity parameter, frequency $\omega_{0}$ or coupling constant
$g$.

On the other hand, if initially we have a disentangled state, for example $\xi=0$, $\phi=0$ in Eq. (\ref{super}),
\begin{equation}
|\psi\rangle_d=|0,1,0,...\rangle_d,
\label{disentangled}
\end{equation}
the concurrence becomes
\begin{equation}
{\cal C}(t,\xi=0,\phi=0)=\frac{1}{2}\left| f_{00}(t)^2-{\rm e}^{-2i\omega_0 t}\right|.
\label{dadoc3}
\end{equation}

For above or other initial conditions 
we can solve the concurrence either
analytically or numerically, depending if cavity radius is infinite
or finite. In this sense, in what follows we consider first the infinite
cavity radius case.

\subsection{Concurrence in free space, $R\to\infty$.}

\begin{figure}[ht]
\includegraphics[scale=0.22]{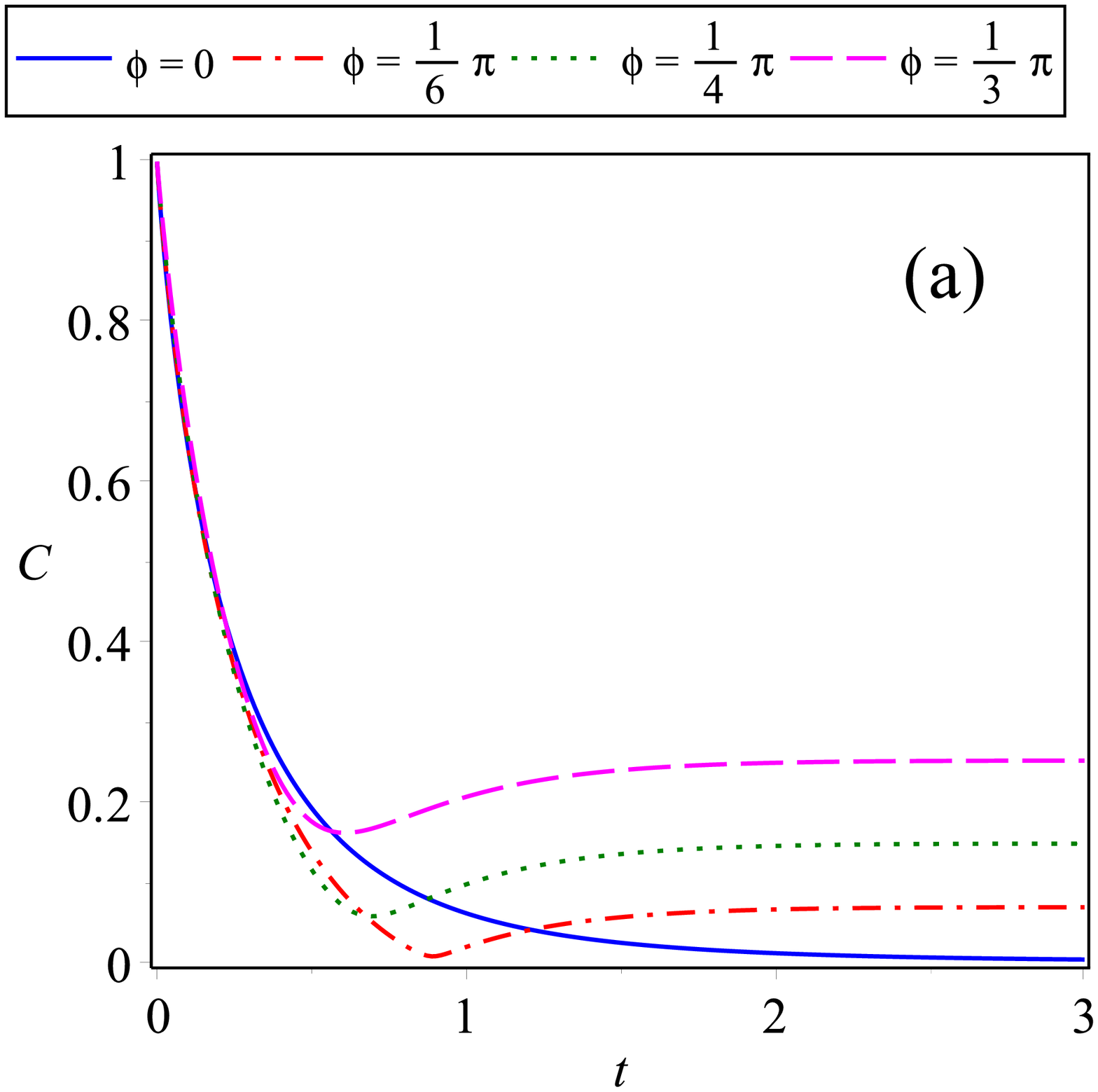}\includegraphics[scale=0.22]{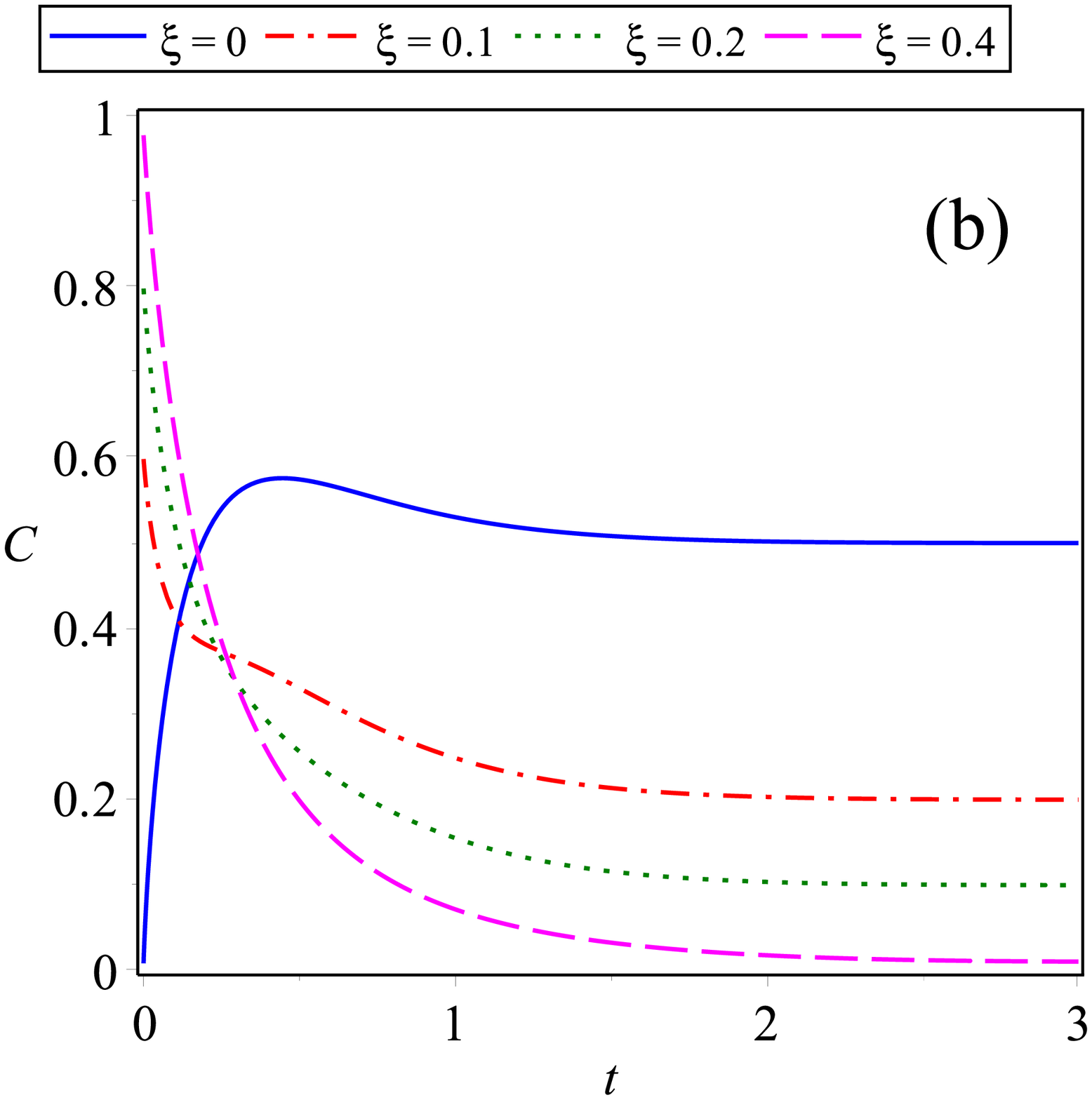}
\caption{(Color online) (a) The time evolution of the concurrence for fixed
$g=\omega_0$, assuming initial state (\ref{super}) $\xi=0.5$,
and for a range of values $\phi$. (b) The time evolution of concurrence
for $g=\omega_0$, assuming initial state (\ref{super})
$\phi=0$, and for a range of values $\xi$. Time is in units of $\omega_0^{-1}.$ }
\label{figR} 
\end{figure}

From Eqs. (\ref{concu1}), (\ref{at}), (\ref{bt}), (\ref{ampli1})
and (\ref{ampli2}), we note that the concurrence, for the two-atom state
(\ref{super}) at time $t$, can be expressed  through the time dependent coefficient
$f_{00}(t)$. For infinite cavity, $R\to\infty$, the coefficient
$f_{00}(t)$ is given by \cite{adolfo1}, \cite{gabrielsolo}, 
\begin{equation}
f_{00}(t)=2g\int_{0}^{\infty}dx\frac{x^{2}\mathrm{e}^{-ixt}}{(x^{2}-\omega_{0}^{2})^{2}+\pi^{2}g^{2}x^{2}}\,.\label{foofree}
\end{equation}
Therefore, for sufficiently large $t\to\infty$, we have $f_{00}(t)\to0$,
thus, the concurrence becomes 
\begin{equation}
{\cal C}(t\to\infty)=\frac{1}{2}-\sqrt{\xi(1-\xi)}\cos\phi\,.\label{concuinfty}
\end{equation}
From above equation we can
note that, for any initial state of the type given by (\ref{super}),
in general the two-atom system is
never disentangled for large $t$ unless  $\sqrt{\xi(1-\xi)}\cos\phi=1/2$.
It is not difficult to
see that above condition is satisfied only if $\xi=0.5$ and $\phi=0$,
\textit{i.e}, the only initial state of the type (\ref{super}) that becomes disentangled for large time $t$
is the symetric state given by Eq. (\ref{symmetric}). For this initial state, in Fig. \ref{figR}-(a),  the
time evolution of the concurrence is depicted in solid line for $g=\omega_0$, where concurrence
decays almost exponentially from 1 to 0. For other initial states, the two-atom system never disentangle,
including an initial non entangled state as the one given by Eq. (\ref{disentangled}). In this case, as
showed in Fig. \ref{figR}-b, solid curve, concurrence increases from 0 to 1/2 according (\ref{concuinfty}).

On the other hand,
at time $t\to\infty$ a maximal entangled state is possible  if
$\sqrt{\xi(1-\xi)}\cos\phi=-1/2$, a relation satisfied only if $\xi=1/2$
and $\phi=\pi$, \textit{i.e}, if the initial state is the antisymmetric  one for
which the system is maximally entangled all the time, Eq. (\ref{dadoc2}).

For other initial states and finite time, $t$, we show in Fig. \ref{figR} the time evolution
of the concurrence for coupling constant $g=\omega_0$. 
In Fig. \ref{figR}-(a) it is considered four initial states with common $\xi=0,5$
and $\phi=0,\pi/6,\pi/4,\pi/3$. In Fig. \ref{figR}-(b), it is considered
four initial states with fixed $\phi=0$ and $\xi=0, 0.1, 0.2, 0.4$.
In all these cases we note a non oscillatory behavior of the concurrence
as a function of time and in agreement with Eq. (\ref{concuinfty}),
the concurrence approaches a definite value at $t\to\infty$. For other values
of the coupling constant $g$ the qualitative behaviour in time will be the same.
If $g>\omega_0$, the assymptotic value of the concurrence will be reached faster than where $g=\omega_0$ and
if $g<\omega_0$ we will have a contrary effect.

From  Fig. \ref{figR}-(a), we note that for some particular initial initial states, for example $\xi=0.5$, $\phi=\pi/6$,  the
concurrence approaches zero for finite time $t$. It is not difficult to  find the condition for a disentangled state
at finite time $t$. From Eq. (\ref{concu1}), concurrence will be zero if $A(t)=0$ or $B(t)=0$,
and using Eqs. (\ref{at}) and (\ref{bt}) we get 
\begin{eqnarray}
&&(2\xi-1)(|f_{00}(t)|^2+1)+2\Re\left(f_{00}(t){\rm e}^{i\omega_0 t}\right)=0,
\label{phi}\\
&& e^{i\phi}=\pm\frac{\sqrt{\xi}}{\sqrt{1-\xi}}\frac{\left(f_{00}(t)+{\rm e}^{-i\omega_0 t}\right)}
{\left(f_{00}(t)-{\rm e}^{-i\omega_0 t}\right)}.
\label{czero}
\end{eqnarray}
Above equations, valid for arbitrary cavity size, must  be used as follows. Given $\xi$ and $\phi$ the
concurrence must be zero at time $t$, if and only if the pair of Eqs. (\ref{phi})-(\ref{czero})  are satisfied simultaneusly. 
Of course for arbitrary values of $\xi$ and $\phi$ above pair of equations are
in general not satisfied for any time $t$. 
 For infinity  cavity size the function $f_{00}(t)$ given by Eq. (\ref{foofree})
is non oscillating in time, consequently concurrence could vanish only for some finite values of $t$ and
for very restricted initial conditions. The situation will be different when we consider in the next subsection
a cavity of finite size, 
where $f_{00}(t)$ is an oscillating function of time. In such a case it will be possible ${\cal C}=0$, many times,
consequently we will have death and revival in time of entanglement for some  initial states.

\subsection{Concurrence for finite cavity radius.}

Solving numerically the Eq. \eqref{eq:Omega}, we can find the concurrence
using the Eq. \eqref{concu1}. Therefore, in what follows, we will
illustrate some plots of concurrence for a variety of situations,
as a function of time $t$.

\begin{figure}[ht]
\includegraphics[scale=0.43]{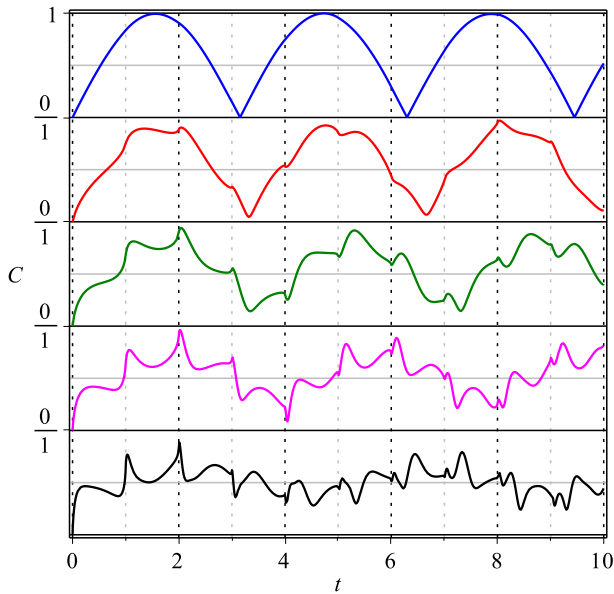} \caption{(Color online) The time evolution of the concurrence for an initial
state (\ref{super}) $\xi=0$ and $\phi=0$, assuming fixed $R\omega_0=1$,
 for a range of values of $g=\{0.01\omega_0,0.1\omega_0,0.5\omega_0,\omega_0,2\omega_0\}$,
from top to bottom, respectively. Time is in units of $\omega_0^{-1}$. }
\label{Fig-conc1} 
\end{figure}

In Fig. \ref{Fig-conc1}, we illustrate the time evolution of concurrence,
assuming fixed parameter cavity radius, $R=\omega_0^{-1}$, and initial
conditions $\xi=0$ and $\phi=0$,  [fully disentangled condition, see Eqs. (\ref{disentangled})- (\ref{dadoc3})],
as a function of time $t$, for a range of values of 
$g=\{0.01\omega_0,0.2\omega_0,0.5\omega_0,\omega_0,2\omega_0\}$
from top to bottom. In the top of Fig. \ref{Fig-conc1} is illustrated
the time evolution curve for $g=0.01\omega_0$, and one can observe the curve
is some what periodic $\mathcal{C}\approx|\cos(\omega_{0}t)|$. In
below the top of Fig. \ref{Fig-conc1}, is depicted for $g=0.1\omega_0$, thus,
the periodicity is definitely wrecked, although the curve still holds
some periodic pattern. For the middle ($g=0.5\omega_0$) and above the below
of Fig. \ref{Fig-conc1}, for fixed $g=\omega_0$, the periodic curve is
strongly modified and  becomes almost a random curve without
any periodicity. In bottom of Fig. \ref{Fig-conc1}, is illustrated
the curve for $g=2\omega_0$, and the concurrence fluctuate in time around
a constant $\mathcal{C}=0.5$, with small short quasi-random oscillations
and no periodicity is observed. From these curves, we can conclude that for fixed
cavity radius, the concurrence oscilates for suficiently weak coupling constant between its
minimum and maximum value, {\it i.e} we have a constant revival and death of
entanglement when the two atom-system evolves in time. When coupling constant is increased, 
such effect is suppressed and the concurrece oscillates around its infinity cavity value for
large $t$ (compare solid line of Fig. \ref{figR}-(b) with the curve at bottom of Fig. \ref{Fig-conc1}).

\begin{figure}[ht]
\includegraphics[scale=0.225]{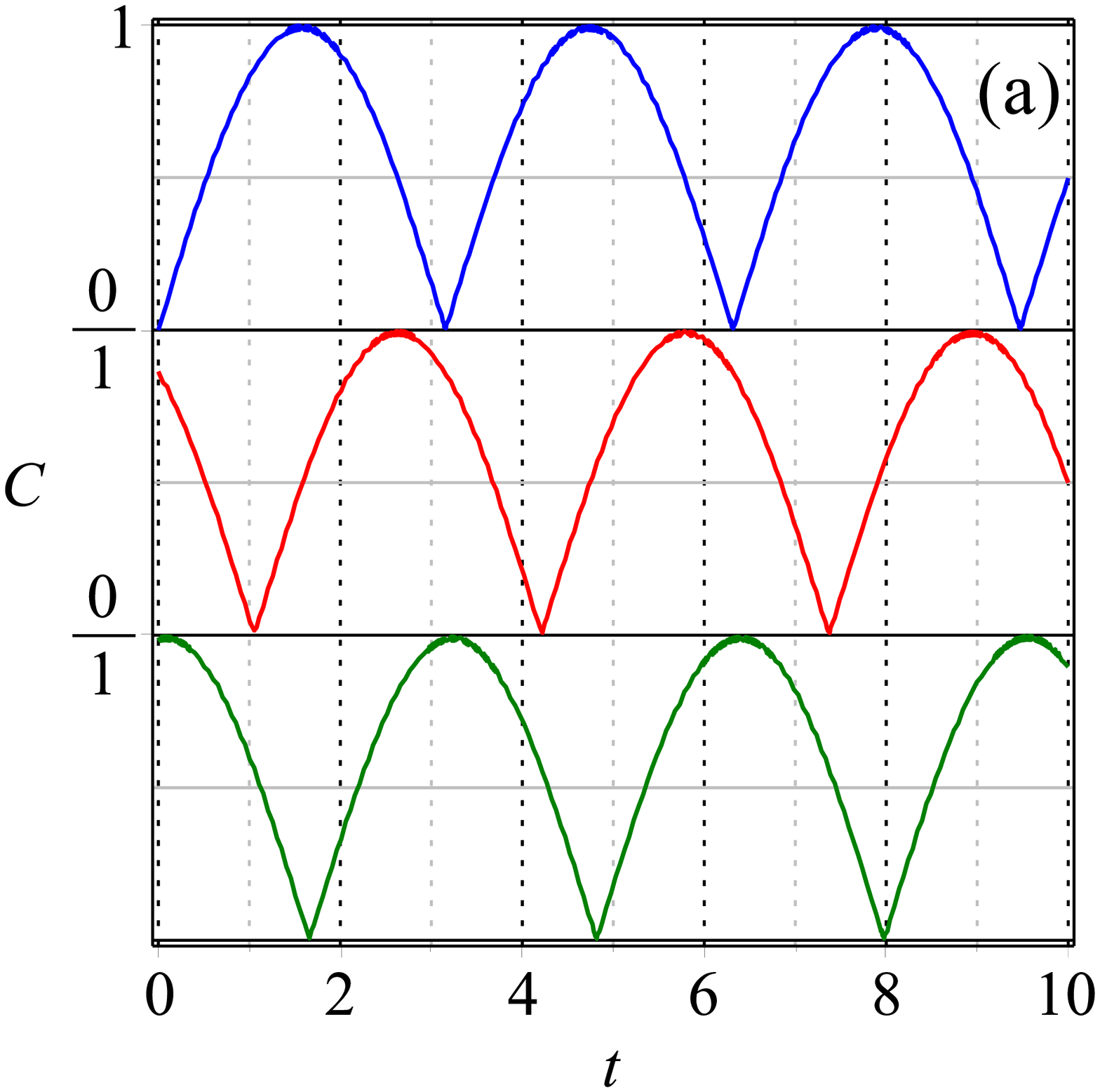}\includegraphics[scale=0.225]{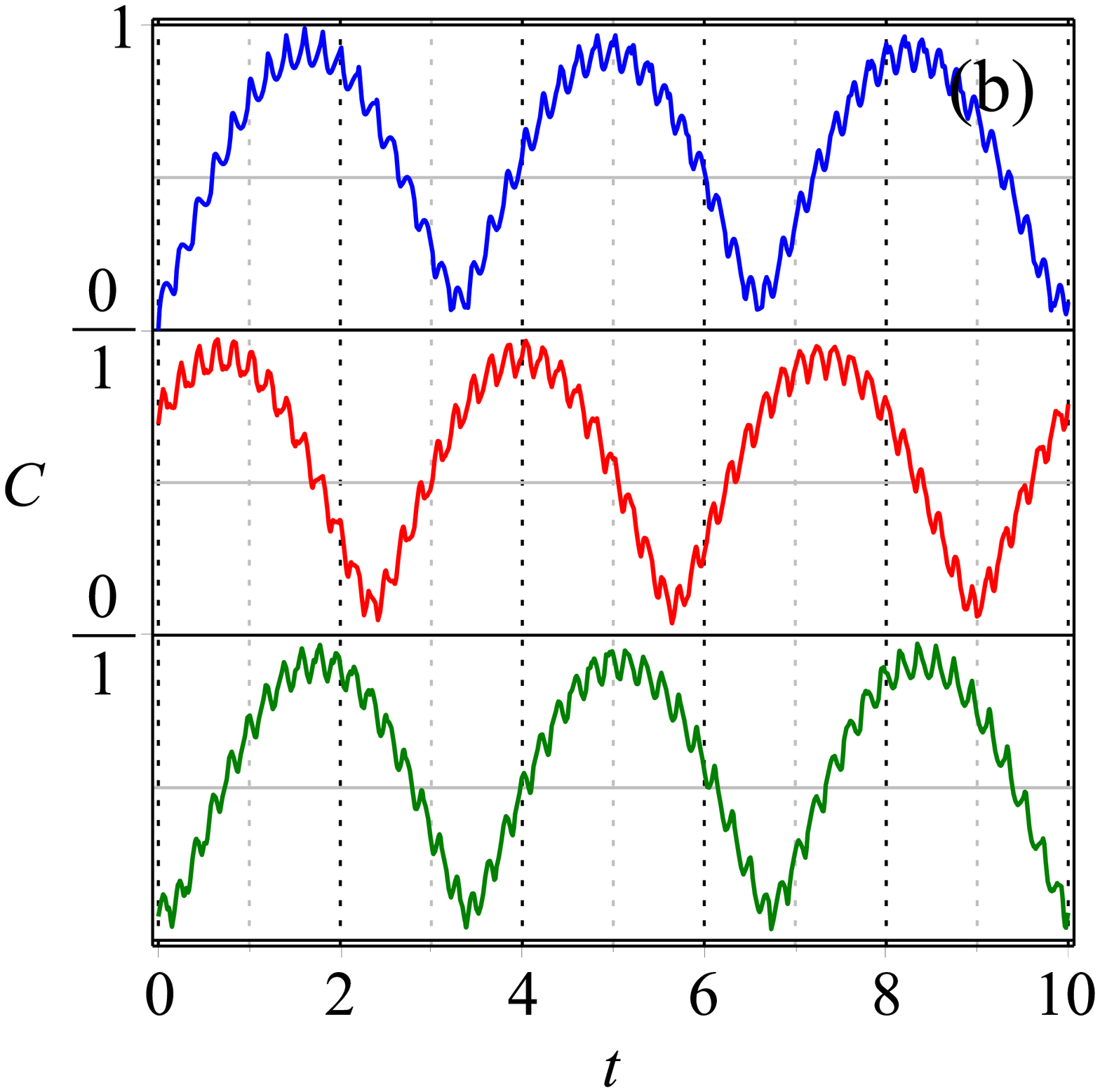}

\includegraphics[scale=0.225]{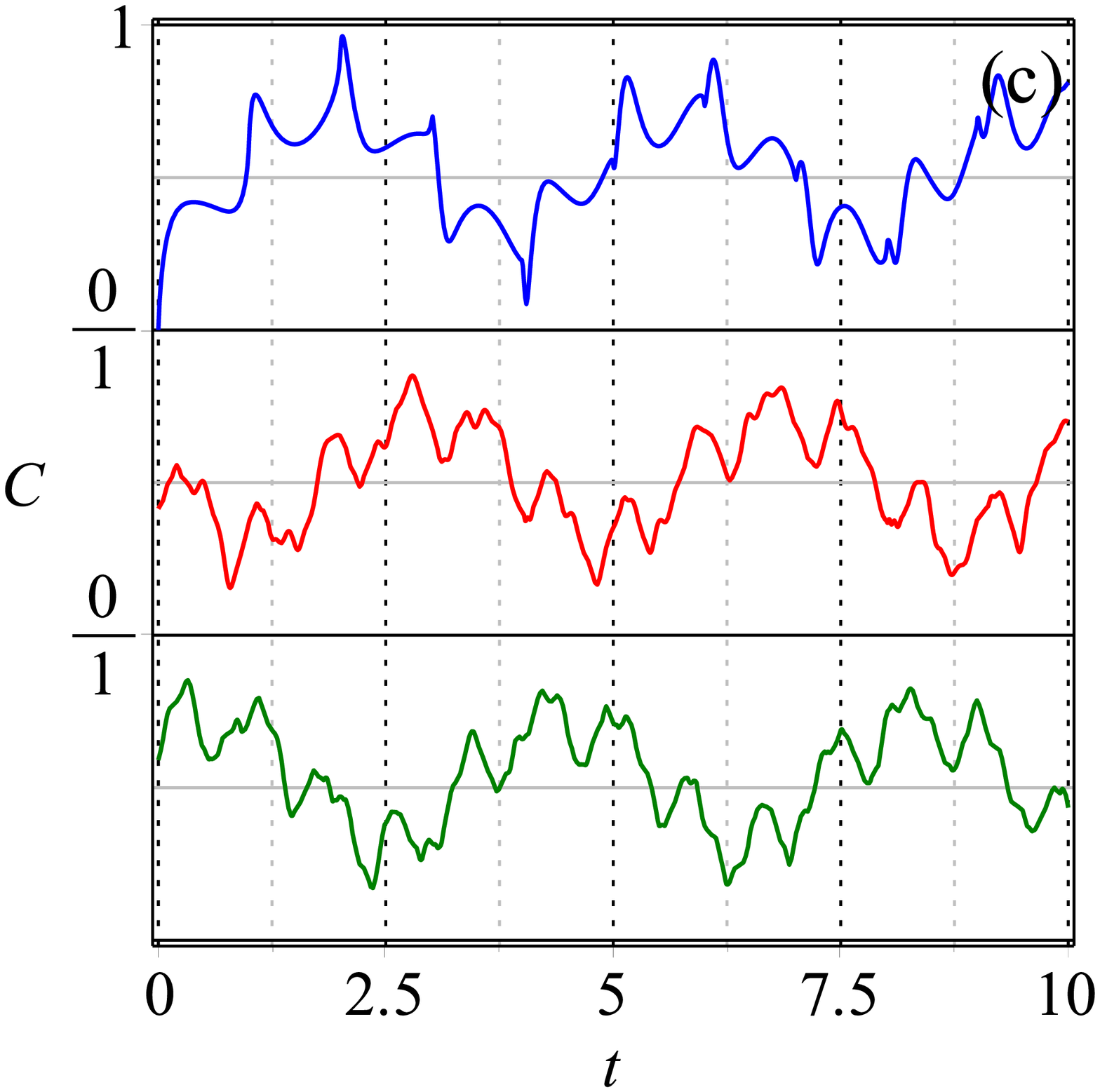}\includegraphics[scale=0.225]{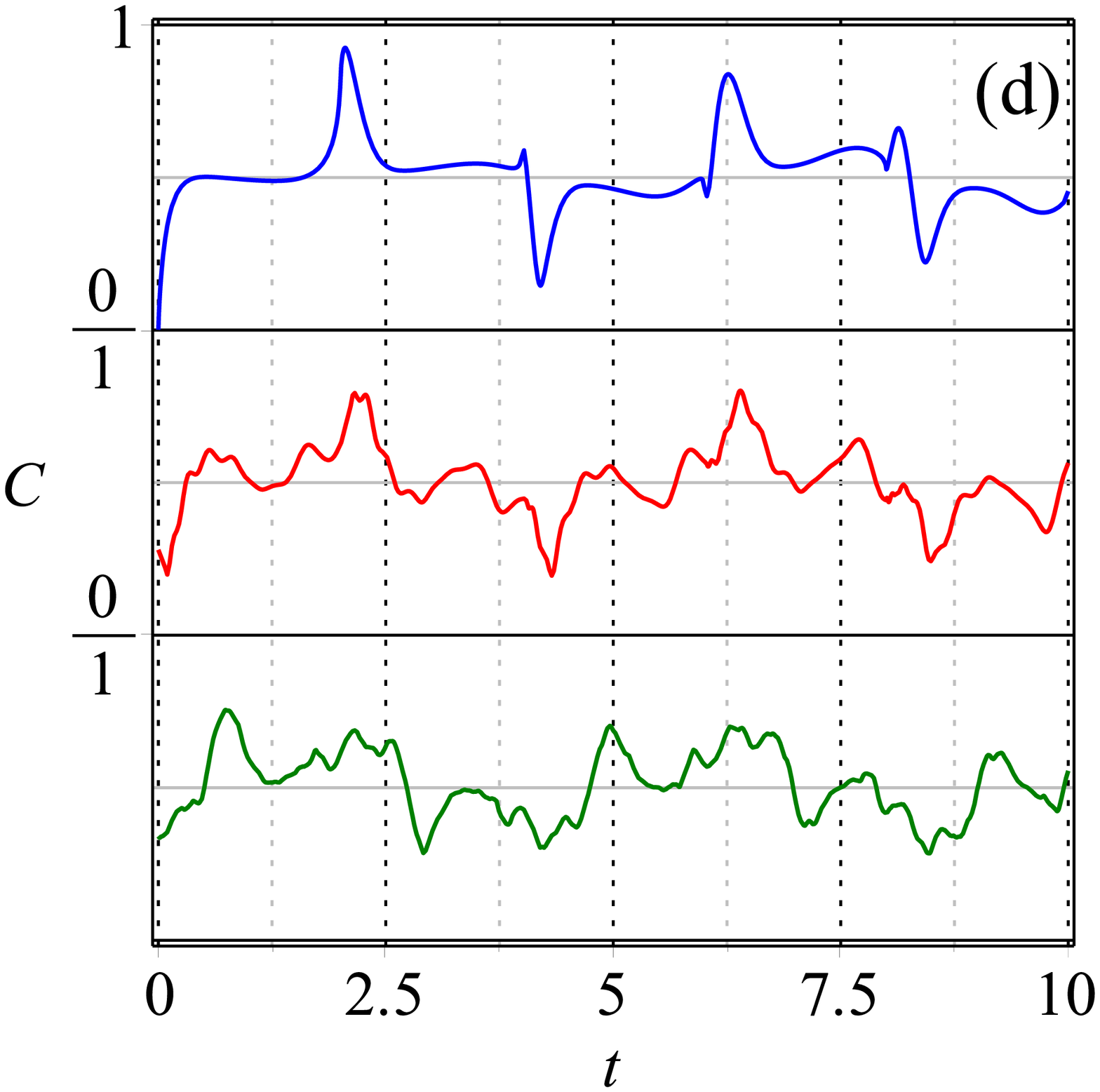}

\caption{(Color online) Concurrence as a function of time, for an initial state
$\xi=0$ and $\phi=0$, and considering fixed $g=\omega_{0}$.
(a) Displays for fixed $R\omega_0=0.01$, top curves shows the time evolution
from $t=0...10$, below top curve shows after elapsed a long time,
for an interval $t=10^{2}...10^{2}+10$ , while bottom curve shows
after elapsed a too long time, for an interval $t=10^{9}...10^{9}+10$.
(b) Displays for fixed $R\omega_0=0.1$, following same criterion as in (a),
for $t=0..10$ (top), for $t=10^{2}..10^{2}+10$ (below top) and for
$t=10^{9}..10^{9}+10$ (bottom). (c) Similarly to (a) and (b) illustrates
for $R\omega_0=0.5$. (d) For $R\omega_0=1$. Time is in units of $\omega_0^{-1}$.}
\label{fig3} 
\end{figure}

In Fig. \ref{fig3}, is illustrated the time evolution of the
concurrence in a cavity, for the same initial state (\ref{disentangled}) above
 for fixed $g=\omega_0$, but now for variable cavity radius $R$. (a)
Displays for fixed $R=0.01\omega_0^{-1}$, top curve shows the time evolution from
$t=0...10$ (in units of $\omega_0^{-1}$), below top curve shows after elapsed a long time, for
an interval $t=10^{2}...10^{2}+10$, while the bottom curve shows
after elapsed a too long time, for an interval $t=10^{9}...10^{9}+10$.
We can observe, the concurrence for small cavity behaves almost periodically,
given in the limiting case by $\mathcal{C}\approx|\cos(\omega_{0}t)|$
this relation is still valid even after elapsed a too long time. (b)
Displays for fixed $R=0.1\omega_0^{-1}$, following the same criterion as in Fig. \ref{fig3}-(a),
$t=0..10$ (top), $t=10^{2}..10^{2}+10$ (below top) and $t=10^{9}..10^{9}+10$
(bottom). Here, we observe the rising of a high-frequency oscillation
with small amplitude, but the low frequency oscillation still follows
a pattern of periodicity $\mathcal{C}\approx|\cos(\omega_{0}t)|$.
In Fig. \ref{fig3}-(c) is illustrated similar to Figs. \ref{fig3}-(a) and
\ref{fig3}-(b) for $R=0.5\omega_0^{-1}$. In this case, the high frequency oscillation
is of order of $\omega_{0}$, that means the combination gives quasi-random
oscillation. For long time one can observe the low frequency oscillation
still remains a pattern of oscillation. We can also show the behavior
of this curve after elapsed a too long time, and still remains a similar
curve compared for small \textit{$t$}. Finally, in Fig. \ref{fig3}-(d),
is illustrated for $R=\omega_0^{-1}$, a cavity  large enough compared to
the case of Fig. \ref{fig3}-(a). In this panel we can observe the low
energy frequency oscillation is vanished and high-frequency oscillation
was also wrecked, after elapsed a short time there is only some well
pronounced peak at $t=2R$ (the time necessary for field quanta to go and come back
from the center to the spherical cavity wall), but as
soon as the time evolves these peaks vanishes and after elapsed a
too long time  disappear definitely. From the curves of Fig. \ref{fig3}, we conclude
that for sufficiently small cavity radius, the concurrence oscillates between its minimum
${\cal C}=0$ and maximum value ${\cal C}=1$, consequelty we will have again a
an almost periodical revival and death of entanglement for two-atom states. As for
large coupling constant, this behaviour will be suppressed for suficiently large cavity
radius, where  concurrence will oscillate around
$\mathcal{C}=0.5$, according to its $R\to\infty$ assymptotic  value, Eq. (\ref{concuinfty}).

\begin{figure}[ht]
\includegraphics[scale=0.22]{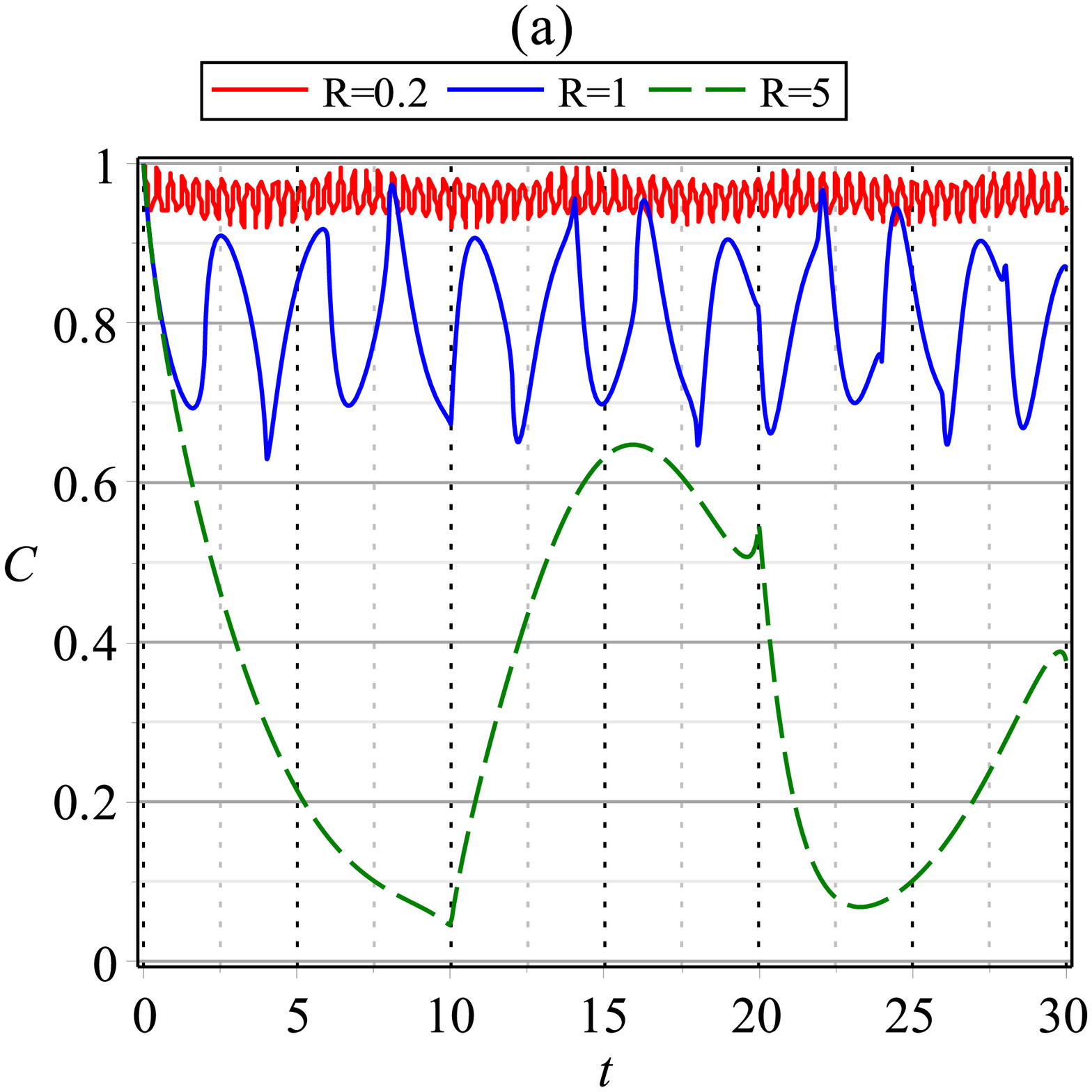}\includegraphics[scale=0.22]{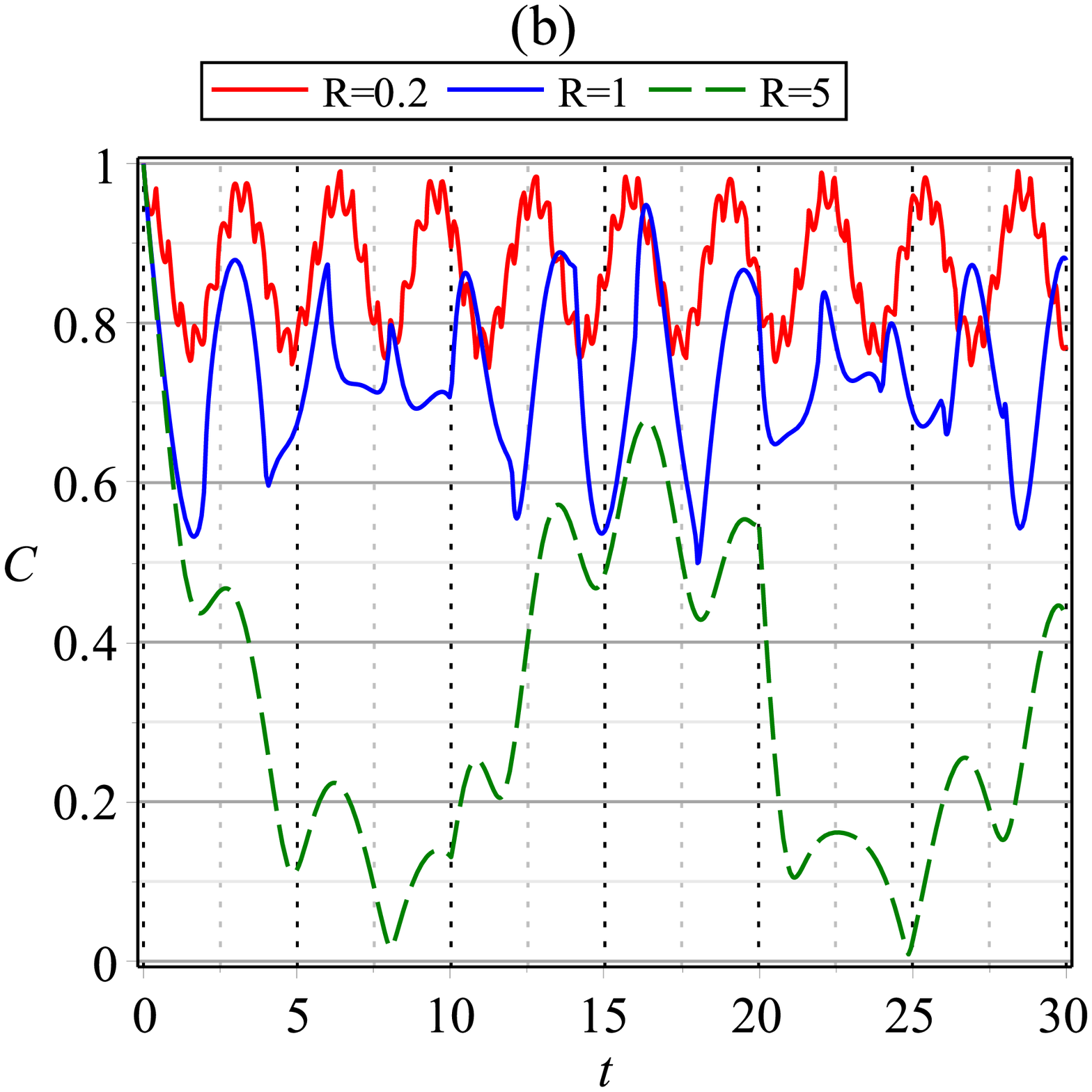}
\caption{(Color online) (a) Concurrence as a function of time, for an initial
state $\xi=0.5$ and $\phi=0$, and considering fixed $g=\omega_{0}$,
and for $R\omega_0=0.2$, $R\omega_0=1$ and $R\omega_0=5$. (b) Concurrence as a function
of time, for an initial state $\xi=0.5$ and $\phi=\frac{\pi}{5}$,
for the same parameters of (a). Time and cavity radius are given in units of $\omega_0^{-1}.$}
\label{fig4} 
\end{figure}

Considering as initial state the maximally symmetric state (\ref{symmetric}), 
 $\xi=0.5$, $\phi=0$  it is illustrated in Fig. \ref{fig4}-(a)  the time evolution of the
concurrence, assuming fixed $g=\omega_{0}$. We considerer $R=0.2\omega_0^{-1}$
(red/gray solid curve), 
$R=\omega_0^{ -1}$ (blue/black solid line)  and $R=5\omega_{0}^{-1}$. We note that for suficiently
small cavity radius the concurrence oscillates below its maximum value $1$ never decreasing appreciably
from this value.
In this case the initial maximally entangled state will remain almost maximally entangled all the time.
Note that this behaviour is dramatically different from the free space case, $R\to\infty$, where the
initial symmetric state is the only one that becomes disentangled for sufficiently large time. According
results above, we expect the same behaviour for sufficiently small coupling constant.
When the cavity radius is increased, the minimum for concurrence decreases, and the frequency of oscillation 
in time increases with $R$. Agai, we expect the same behaviour for increasing coupling constant $g$.
Finally in Fig. \ref{fig4}-(b),
is depicted the time evolution for the initial state $\xi=0.5$, $\phi=\pi/6$,
 assuming the same parameters as in Fig. \ref{fig4}-(a), where we observe qualitativelly the same
behaviour as in above case.

\section{conclusions}

In this work, we considered the study of the entanglement of a two-atom
system driven by  the vacuum quantum field inside a spherical cavity, using the dressed
coordinates approach. The time evolution entanglement was analyzed,
for a variety of initial states composed as a superposition of atomic
states, Eq. (\ref{super}). For  free space, $R\to\infty$, we find that concurrence approaches
assymptotically  a fixed value at sufficiently large time. From this
result we concluded that with exception of the initial symmetric maximally
entangled state, $\xi=0.5,\phi=0$ all  the other states become or remain entangled for sufficiently large
time. For the initial state  $\xi=0.5,\phi=0$
the concurrence decreases,
as a function of time, almost exponentially from its maximum value
to zero. Also we showed that there are  very restricted initial states, who satisfies Eqs. (\ref{phi})-(\ref{czero}),
 for which
the two-atom state becomes untangled for a finite value of time. Anyway,
for large $t$ also those states become or remain entangled.

For finite cavity radius, we found that
concurrence in general behaves almost as a periodic function of time, for  sufficiently small cavity  radius
or coupling constant. 
For the initial disentangled state $\xi=0$, $\phi=0$, the concurrence oscilates almost
periodically for small cavity radius or small coupling constant, between
${\cal C}=0$ and ${\cal C}=1$. When $R$ or $g$ are increased the
aforementioned oscillation is supressed and concurrence approaches
its infinity cavity size value, ${\cal C }=1/2$, see Figs. \ref{Fig-conc1} and \ref{fig3}.  
On the other hand for the initial
symmetric maximally entangled state, $\xi=0.5$, $\phi=0$, the concurrence
is always close to ${\cal C}=1$, when the cavity radius or the
coupling constant are sufficiently small and for larger cavity radius
or coupling constant, the concurrence decays considerably from its
initial value, leading to a small concurrence at $t=2R$, see Fig. \ref{fig4}-(a).

In general, when the system is initially maximally entangled (disentangled)
after a time, proportional to the cavity radius, had elapsed the
system becomes again  strongly entangled (disentangled),
particularly during the first oscillations, later this phenomenon
could be wrecked depending on the initial condition. Another interesting
result we found is the behavior of the concurrence after too long
time in the future, for large ($R\neq \infty$) cavity size, apparently there is no pattern resembling some
periodicity. 

As pointed out earlier, for fre space, $R\to\infty$, there are some initial entangled states that become  disentangled 
for finite values of time that satisfy Eqs. (\ref{phi})-(\ref{czero}). As showed such condition is valid
for any cavity radius and since for finite cavity radius, $f_{00}(t)$ has an almost periodic behavior for
suficiently small cavity or coupling constant, then we expect and oscillatory behaviour for concurrence
between its initial value and ${\cal C}=0$, {\it i.e}, we will have sudden death and revival for those
initial conditions, according to the one found in Ref. \cite{paz}, at zero temperature. We have
to remark here that althought the hamiltonian given by Eq. (\ref{hamiltonian}) is the same used in
Ref. \cite{paz}, physically the system we treat here is very different, since if written in terms
of dressed coordinates (that we take as the physical coordinates) hamiltonian (\ref{hamiltonian}) will be very different to the one
of Ref. \cite{paz}.  

Also, independent of the cavity radius or other parameters,  we found that the initial 
anti-symmetric maximally entangled state, corresponding to $\xi=0.5, \phi=\pi$, remains
fully entangled all the time. 
The reason for this behavior is because
such initial state is an eigenfunction of the total system and consequently
remains stable. At this point we have to mention that a similar result has been found recently
in Ref. \cite{nami}, where the authors showed at first order in perturbation theory,
that an initial anti-symmetric two-atom state remains stable when the spatial separation between the atoms
is zero. 

We call atention to the fact that in Ref. \cite{malbouissonx} the authors considered the same model here. However 
the authors considered the time evolution of a superposition of dressed states related to the center of mass 
and relative coordinates. Here we have defined, dressed coordinates and states for the atoms $A$ and $B$, 
consequently our results and conclusions are quit different. 

In this work we considered the case in which the field is initially in the vacuum
state and this situation is physically equivalent to one in which
the field system is at zero temperature. Then, all above conclusions
are valid in such situation
and our results are  similar to the found in Ref. \cite{paz},  for $T=0$ with one exception below. We have
found in general two phases, suden death and revival (SDR) and non suden death (NSD) depending on the
initial state, cavity radius and coupling constant. But when considering as initial state the symmetric maximally entangled one
we found sudden death (SD), at large time in free space, $R\to\infty$.
A natural extension of this work is to consider  finite temperature effects, where the dynamical
entanglement could show a SD phase for another initial states and perhaps for finite cavity size. 
We expect to report about that elsewhere.

\section*{Acknowledgments}

This work was partially supported by Brazilian agencies CNPq and FAPEMIG.

\end{document}